\documentclass[12pt]{article}
\usepackage{epsf}
\usepackage{amsmath,amssymb}
\usepackage{epstopdf}
\usepackage{graphicx}
\usepackage{cite}

\usepackage{comment}

\begin{document}

\newcommand{\lsim}{\stackrel{<}{_\sim}}
\newcommand{\gsim}{\stackrel{>}{_\sim}}

\newcommand{\rem}[1]{{$\spadesuit$\bf #1$\spadesuit$}}

\renewcommand{\thefootnote}{\fnsymbol{footnote}}
\setcounter{footnote}{0}

\begin{titlepage}

\def\thefootnote{\fnsymbol{footnote}}

\begin{center}

\hfill UT-16-24\\
\hfill IPMU-16-0082\\
\hfill TU-1024\\
\hfill June, 2016

\vskip .5in

{\Large \bf 

  Upper Bound on the Gluino Mass
  \\
  in Supersymmetric Models with Extra Matters
  \\

}

\vskip .5in

{\large 
  Takeo Moroi$^{(a,b)}$,
  Tsutomu T. Yanagida$^{(b)}$ and
  Norimi Yokozaki$^{(c)}$
}

\vskip 0.25in

\vskip 0.25in

$^{(a)}${\it Department of Physics, University of Tokyo, 
Tokyo 113-0033, Japan}
\vspace{0.3cm}

$^{(b)}${\it Kavli Institute for the Physics and Mathematics of the Universe 
(Kavli IPMU), \\
University of Tokyo, Kashiwa 277--8583, Japan}

\vspace{0.3cm}
$^{(c)}${\it Department of Physics, Tohoku University,
Sendai 980-8578, Japan}

\end{center}
\vskip .5in

\begin{abstract}

  We discuss the upper bound on the gluino mass in supersymmetric
  models with vector-like extra matters.  In order to realize the
  observed Higgs mass of $125\ {\rm GeV}$, the gluino mass is bounded
  from above in supersymmetric models.  With the existence of the
  vector-like extra matters at around TeV, we show that such an upper
  bound on the gluino mass is significantly reduced compared to the
  case of minimal supersymmetric standard model.  This is due to the
  fact that radiatively generated stop masses as well the stop
  trilinear coupling are enhanced in the presence of the vector-like
  multiplets. In a wide range of parameter space of the model with
  extra matters, 
  particularly with sizable $\tan\beta$ (which is the ratio of the vacuum expectation values of the two Higgs
  bosons), the gluino is required to be lighter than $\sim 3\ {\rm TeV}$, which is likely to be within the reach of forthcoming LHC experiment.

\end{abstract}

\end{titlepage}

\renewcommand{\thepage}{\arabic{page}}
\setcounter{page}{1}
\renewcommand{\thefootnote}{\#\arabic{footnote}}
\setcounter{footnote}{0}

\section{Introduction}

Although the low-energy supersymmetry (SUSY) is attractive from the
points of view of, for example, naturalness, gauge coupling
unification, dark matter, and so on, to which the standard model (SM)
has no clue, no signal of the SUSY particles has been observed yet.
Thus, one of the important questions in the study of models with
low-energy SUSY is the scale of SUSY particles.

It is well-known that the observed Higgs mass of $\sim 125\ {\rm GeV}$
\cite{Aad:2015zhl} gives information about the mass scale of SUSY
particles (in particular, stops).  The Higgs mass is enhanced by
radiative corrections when the stop masses are much larger than the
electroweak scale \cite{Okada:1990vk, Okada:1990gg, Ellis:1990nz,
  Ellis:1991zd, Haber:1990aw}.  Thus, the stop masses are bounded from
above in order not to push up the Higgs mass too much; the stop masses
are required to be smaller than $10^4-10^5\ {\rm GeV}$ as far as
$\tan\beta$, which is the ratio of the vacuum expectation value of the
up-type Higgs to that of the down-type Higgs, is larger than a few.
(For the recent study of such an upper bound, see, for example,
\cite{Bagnaschi:2014rsa}.)  Then, too large gluino masses are also
disfavored because, via renormalization group (RG) effects, it results
in stops which are too heavy to make the Higgs mass consistent with
the observed value.  Such an upper bound on the gluino mass is
important for the future collider experiments, in particular, for the
LHC Run-2, in order to discover and to study models with low energy
SUSY. The purpose of this letter is to investigate how such an upper
bound on the gluino mass depends on the particle content of the model.

We pay particular attention to SUSY models with extra vector-like
chiral multiplets which have SM gauge quantum numbers.  In these days,
such extra vector-like matters are particularly motivated from the
excess of the diphoton events observed by the LHC
\cite{ATLAS-CONF-2015-081, ATLAS-CONF-2016-018, CMS:2015dxe,
  CMS:2016owr}.  The most popular idea to explain the diphoton excess
is to introduce a scalar boson $\Phi$ with which the LHC diphoton excess
can be due to the process $gg\rightarrow\Phi\rightarrow\gamma\gamma$.
In such a class of scenarios, vector-like particles which interact
with $\Phi$ are necessary to make $\Phi$ being coupled to the SM gauge
bosons.  Indeed, it has been shown that the LHC diphoton excess are
well explained in SUSY models with vector-like chiral multiplets
\cite{Angelescu:2015uiz, Buttazzo:2015txu, Franceschini:2015kwy, Ellis:2015oso,
  Patel:2015ulo, Craig:2015lra, Hall:2015xds, Tang:2015eko,
  Wang:2015omi, Palti:2016kew, Chao:2016mtn, Dutta:2016jqn,
  Djouadi:2016eyy, King:2016wep,
  Han:2016fli,Barbieri:2016cnt,Nilles:2016bjl,Hall:2016swn,Cohen:2016kuf}.
Assuming a perturbative gauge coupling unification at the GUT scale of
$\sim 10^{16}\ {\rm GeV}$,\footnote
{For the perturbativity bounds on models with extra matters, see
\cite{Bae:2016xni}.}
three or four copies of the vector-like multiplets, which transform
${\bf 5}$ and $\bar{\bf 5}$ in $SU(5)$ gauge group, are suggested, and
their masses need to be around or less than 1\,TeV.  In addition, the
vector-like chiral multiplets are also motivated in models with
non-anomalous discrete $R$-symmetry \cite{Kurosawa:2001iq,
  Asano:2011zt}.

With extra vector-like chiral multiplets, the RG evolutions of the
coupling constants and mass parameters of the SUSY models drastically
change compared to those in the minimal SUSY SM (MSSM).  Consequently,
as we will see, the upper bound on the gluino mass becomes
significantly reduced if there exist extra vector-like chiral
multiplets.  Such an effect has been discussed in gaugino mediation
model \cite{Moroi:2012kg} and in the light of recent diphoton excess
at the LHC \cite{Han:2016fli}.  

In this letter, we study the upper bound on the gluino mass in SUSY
models with extra vector-like matters, assuming more general framework
of SUSY breaking.  We extend the previous analysis and derive the
upper bound on the gluino mass.  We will show that the bound on the
gluino mass is generically reduced with the addition of extra matters.
The upper bound becomes lower as the number of extra matters
increases, and the bound can be as low as a few TeV which is within
the reach of the LHC Run-2 experiment.

\section{Enhanced Higgs boson mass and gluino mass}

We first explain how the upper bound on the gluino mass is reduced in
models with extra vector-like multiplets.  To make our discussion
concrete, we consider models with extra chiral multiplets which can be
embedded into complete $SU(5)$ fundamental or anti-fundamental
representation as $\bar{\bf 5}_i=(\bar D_i', L_i') $ and ${\bf
  5}_i=(D_i', \bar L_i')$; we introduce $N_5$ copies of ${\bf 5}$ and
$\bar{\bf 5}$ with $i=1 \dots N_5$.\footnote
{Our results are qualitatively unchanged even if the vector-like
  matters are embedded into other representations of $SU(5)$, as far
  as the parameter $N_5$ is properly interpreted.  For the case with
  $N_{10}$ copies of $\overline{\bf 10}$ and ${\bf 10}$
  representations, for example, $N_5$ should be replaced by
  $3N_{10}$.}
Then, the superpotential is given by
\begin{eqnarray}
W = W_{\rm MSSM} + M_V (\bar D'_i D'_i +\bar L'_i L'_i),
\end{eqnarray}
where $W_{\rm MSSM}$ is a superpotential of the MSSM and $M_V$ is the
common masses for vector-like matter fields.\footnote
{Due to the RG runnings, the SUSY invariant masses for $D'$ and $L'$
  should differ even if they are unified at the GUT scale.  Such an
  effect is, however, unimportant for our following discussion, and we
  neglect the mass difference among the extra matters.}
Hereafter, $M_V$ (=$M_{D'}$=$M_{L'}$) is taken to be $\sim 1\ {\rm
  TeV}$, while $N_5=3$ and $4$, which are suggested by, for example,
the diphoton excess observed by the LHC \cite{Buttazzo:2015txu,
  Franceschini:2015kwy, Ellis:2015oso, Patel:2015ulo, Craig:2015lra,
  Hall:2015xds, Tang:2015eko, Wang:2015omi, Palti:2016kew,
  Chao:2016mtn, Dutta:2016jqn, Djouadi:2016eyy, King:2016wep,
  Han:2016fli,Barbieri:2016cnt,Nilles:2016bjl,Hall:2016swn,Cohen:2016kuf}.\footnote
{The case of $N_5=3$ is particularly interesting, since it may be
  embedded into an $E_6$ GUT \cite{Cai:2016ymq}.}

In order to see how the upper bound on the gluino mass is derived, it
is instructive to see the leading one-loop correction to the Higgs
mass.  Assuming that the left- and right-handed stop masses are almost
degenerate, the Higgs boson mass with the leading one-loop corrections
in the decoupling limit is estimated as~\cite{Okada:1990vk,
  Okada:1990gg, Ellis:1990nz, Ellis:1991zd, Haber:1990aw}
\begin{eqnarray}
  m_h^2 \simeq m_Z^2 \cos^2 2\beta
  + \frac{3}{4\pi^2} \frac{m_t^4}{v^2}
  \left[
    \ln  \frac{M_{\tilde{t}}^2}{m_t^2}  
    + \frac{|X_t|^2}{M_{\tilde{t}}^2}
    \left(1 - \frac{|X_t|^2}{12 M_{\tilde{t}}^2} \right)
  \right],
  \label{mh}
\end{eqnarray}
where $m_Z$ is the $Z$-boson mass, $m_t$ is the top mass,
$M_{\tilde{t}}$ is the stop mass, $v=174.1$ GeV is the vacuum
expectation value of the Higgs boson, and $X_t= A_t - \mu/\tan\beta$
(with $A_t$ being the trilinear coupling of stops normalized by the
top Yukawa coupling constant $y_t$, and $\mu$ being the Higgsino
mass).  Notice that the first term in the square bracket of Eq.\
\eqref{mh} is the effect of the RG running of the quartic SM Higgs
coupling constant from the mass scale of the SUSY particles to the
electroweak scale, while the second one is the threshold correction at
the mass scale of the SUSY particles.  The Higgs mass becomes larger
as $M_{\tilde{t}}$ or $X_t$ increases (as far as $X_t\lesssim
\sqrt{6}$).  Thus, in order to realize the observed value of the Higgs
mass, $m_h \simeq 125$ GeV, there is an upper-bound on $M_{\tilde{t}}$
and $X_t$.  Importantly, the stop masses and the $A_t$ parameter are
enhanced with larger value of the gluino mass because of the RG
runnings from a high scale to the mass scale of SUSY particles.
Consequently, with boundary conditions on the MSSM parameters given at
a high scale, we obtain the upper bound on the gluino mass to have
$m_h\simeq 125\ {\rm GeV}$.  Hereafter, we assume that the MSSM is
valid up to the GUT scale $M_{\rm GUT}\sim 10^{16}\ {\rm GeV}$ and
derive such an upper bound.

Now we consider how the existence of the extra matters affects the
upper bound on the gluino mass by using one-loop RG equations (RGEs),
although two-loop RGEs are used for our numerical calculation in the
next section.  With $N_5$ pairs of the vector-like multiplets, RGEs of
gauge coupling constants at the one-loop level are
\begin{eqnarray}
  \frac{d g_i}{d \ln \mu_R} = \frac{b_i}{16\pi^2} g_i^3,
\end{eqnarray}
where $\mu_R$ is a renormalization scale; $g_1$, $g_2$ and $g_3$ are
gauge coupling constants of $U(1)_Y$ (in $SU(5)$ GUT normalization),
$SU(2)_L$ and $SU(3)_C$, respectively.  In addition,
$(b_1,b_2,b_3)=(33/5+N_5, 1+N_5, -3 + N_5)$.  For $M_V \lesssim
1$\,TeV, $N_5 \lesssim 4$ needs to be satisfied under the condition
that the coupling constants remain perturbative up to the GUT scale
($\sim 10^{16}$\,GeV). As one can see, for $N_5\gtrsim 3$, $g_3$ is
not asymptotically free.  One-loop RGEs of gaugino masses are
\begin{eqnarray}
\frac{d M_i}{d \ln \mu_R} = \frac{b_i}{8\pi^2} g_i^2 M_i,
\end{eqnarray}
where $M_1$, $M_2$ and $M_3$ are the Bino, Wino and gluino mass,
respectively.  (Hereafter, we use the convention in which $M_3$ is
real and positive.)  The ratio $M_i/g_i^2$ is constant at the one-loop
level and, with $M_i$ at the mass scale of the SUSY particles being
fixed, the gaugino masses at higher scale are more enhanced with
larger value of $N_5$.  In particular, for $N_5 \gtrsim 3$, the gluino
mass, whose RG effects on $A_t$ and the stop masses are important,
become larger as the RG scale increases.\footnote
{For $N_5=3$, the one-loop beta-function vanishes and $M_3$ is
  constant, but $|M_3|$ becomes larger at the high energy scale due to
  two-loop effects.}
In other words, even if $|M_3|$ is large at the GUT scale, the
low-energy value of $|M_3|$ is small especially for
$N_5=4$~\cite{Han:2016fli}.

With the enhancement of the gluino mass, the RG effect on the $A_t$
parameter becomes larger.  This can be easily understood from the 
RGE of the $A_t$ parameter; at the one-loop level,
\begin{eqnarray}
  \frac{d A_t}{d \ln \mu_R} = \frac{1}{16\pi}
  \left[
    \frac{32}{3} g_3^2 M_3 + 6 y_t^2 A_t + \cdots
  \right],
\end{eqnarray}
where we show only the terms depending on $SU(3)_C$ gauge coupling
constant or the top Yukawa coupling constant.  One can see that the
$A_t$ parameter is generated by the RG effect using the gluino mass as
a source, and the low-energy value of $|A_t|$ is likely to become
larger as $|M_3|$ increases.

More quantitative discussion about the enhancement of the $A_t$
parameter is also possible.  Solving RGEs, the $A_t$ parameter at the
mass scale of the SUSY particles, denoted as $m_{\mathcal{S}}$, can be
parametrized as
\begin{align}
  A_t (m_{\mathcal{S}}) \simeq
  \left\{
    \begin{array}{c}
      -0.77  \\
      -1.84  \\
      -5.18
    \end{array}
  \right\} M_3 (m_{\mathcal{S}})
  + 
  \left\{
    \begin{array}{c}
      0.39  \\
      0.47  \\
      0.36
    \end{array}
  \right\} A_0,
  \label{At(fit)}
\end{align}
where the numbers in the curly brackets are the coefficients for the
cases of the MSSM (i.e., $N_5=0$), $N_5=3$, and $N_5=4$, from the top
to the bottom, which are evaluated by using two-loop RGEs with
$m_{\mathcal{S}}=3.5\ {\rm TeV}$, and $A_0\equiv A_t(M_{\rm inp})$
with $M_{\rm inp}$ being the scale where the boundary conditions for
the SUSY breaking parameters are set.  (In our numerical calculations,
we take $M_{\rm inp}=10^{16}\ {\rm GeV}$.)  In deriving Eq.\
\eqref{At(fit)} (as well as Eqs.\ \eqref{mQ3(fit)} and
\eqref{mU3(fit)}), we have taken $\tan\beta=10$, and, for simplicity,
we have assumed that (i) the gaugino masses obey the GUT relation,
(ii) the SUSY breaking scalar masses are universal at $M_{\rm inp}$,
and (iii) all the trilinear scalar coupling constants are
proportional to corresponding Yukawa coupling constant (with the
proportionality factor $A_0$) at $M_{\rm inp}$.  We can see that the
coefficient of the $M_3$ term becomes larger as $N_5$ increases.

Similarly, we can discuss how the SUSY breaking stop mass parameters
behave.  Assuming the universality of the scalar masses at
$\mu_R=M_{\rm inp}$,
\begin{align}
  m_{Q_3}^2 (m_{\mathcal{S}}) \simeq &\,
  \left\{
    \begin{array}{c}
      0.68  \\
      2.36  \\
      9.89
    \end{array}
  \right\} M_3^2 (m_{\mathcal{S}})
  + \left\{ \begin{array}{c}
      0.05  \\
      0.21  \\
      1.06
    \end{array}
  \right\} M_3 (m_{\mathcal{S}}) A_0
  + \left\{ \begin{array}{c}
      -0.04  \\
      -0.04  \\
      -0.06
    \end{array}
  \right\} A_0^2
  + \left\{ \begin{array}{c}
      0.66   \\
      0.61   \\
      0.53  
    \end{array}
  \right\} \tilde{m}^2,
  \label{mQ3(fit)}
  \\[7pt]
  m_{\bar U_3}^2 (m_{\mathcal{S}}) \simeq &\,
  \left\{
    \begin{array}{c}
      0.50  \\
      1.39  \\
      3.25
    \end{array}
  \right\} M_3^2 (m_{\mathcal{S}})
  + \left\{ \begin{array}{c}
      0.10  \\
      0.37  \\
      1.30
    \end{array}
  \right\} M_3 (m_{\mathcal{S}}) A_0
  + \left\{ \begin{array}{c}
      -0.08  \\
      -0.08  \\
      -0.05
    \end{array}
  \right\} A_0^2
  + \left\{ \begin{array}{c}
      0.37   \\
      0.40   \\
      0.32   
    \end{array}
  \right\} \tilde{m}^2,
  \label{mU3(fit)}
\end{align}
where $m_{Q_3}$ and $m_{\bar U_3}$ are soft masses of the left-handed
stop and right-handed stop, respectively, and $\tilde{m}$ is the
universal scalar mass.  The coefficients of the $M_3^2$ terms become
significantly enhanced with larger value of $N_5$.  Thus, with the
increase of $N_5$, the stop masses becomes larger with fixed value of
$M_3^2 (m_{\mathcal{S}})$, as far as there is no accidental
cancellation.  In addition, Eqs.\ \eqref{At(fit)}, \eqref{mQ3(fit)} and
\eqref{mU3(fit)} suggest that, for $N_5=3$ and $4$, $A_t
(m_{\mathcal{S}})$, $m_{Q_3}^2 (m_{\mathcal{S}})$, and $m_{\bar U_3}^2
(m_{\mathcal{S}})$ are primarily determined by the gaugino mass if
$M_3$, $A_0$, and $\tilde{m}$ are of the same size.  We can see that,
in such a case, the trilinear coupling constant $A_t$ is more enhanced
than the stop masses, $m_{Q_3}$ and $m_{\bar U_3}$, which makes the
threshold correction to the Higgs mass larger (see Eq.\ \eqref{mh}).

Based on the above discussion, we have seen that the inclusion of the
extra vector-like matter pushes up the Higgs mass for a fixed value of
the gluino mass.  Thus, the relevant value of the gluino mass
realizing the observed Higgs mass becomes lower as the number of extra
matter increases.  In the next section, we will see that this is
really the case, and derived the upper bound on the gluino mass with
more detailed analysis of the RG effects.

\section{Numerical results}

Now, we evaluate the upper-bound on the gluino mass by numerically
solving two-loop RGEs.  For our numerical calculation, we take
$m_t({\rm pole})=173.34$\,GeV~\cite{topmass} and
$\alpha_s(m_Z)=0.1185$~\cite{PDG}.

\subsection{The MSSM results}

\begin{figure}[!t]
\begin{center}
\includegraphics[scale=1.05]{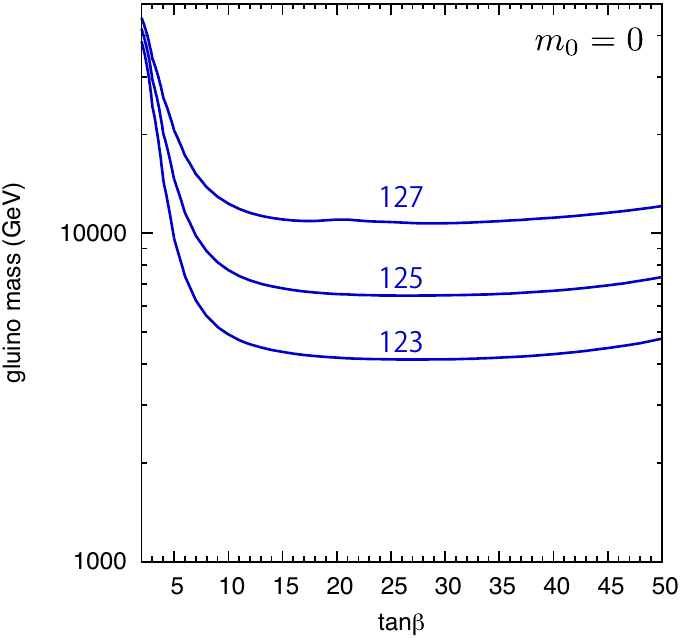}
\includegraphics[scale=1.05]{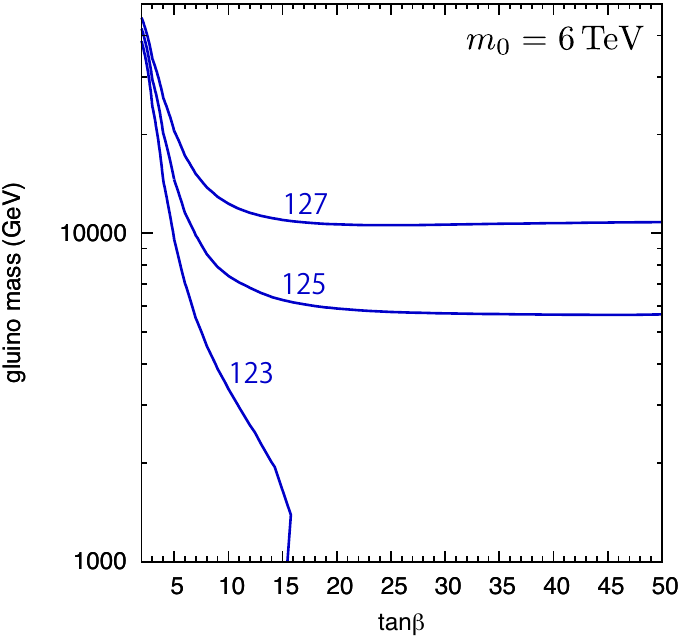}
\caption{
Contours of the Higgs boson mass in the MSSM. 
We take $m_{H_u}=m_{H_d}=0$, $\mu>0$ and $A_0=0$. The other scalar masses are set to be universal value $m_0$.
The solid lines correspond to $m_h=(127,125,123)$\,GeV from top to bottom.
Here, $m_t({\rm pole})=173.34$\,GeV and $\alpha_s(m_Z)=0.1185$.
}
\label{fig:mh_mssm}
\end{center}
\end{figure}
%%%%%%%%%

For the sake of comparison, we first show the upper-bound without
vector-like multiplets, i.e. in the case of the MSSM. In the
calculation, we take
\begin{eqnarray}
M_1=M_2=M_3=M_{1/2}, \ m_{H_u}=m_{H_d}=0, \ A_0=0, \ \mu>0 \ {\rm at} \ M_{\rm inp}, \label{eq:cond1}
\end{eqnarray}
with $M_{\rm inp}=10^{16}$\,GeV; $A_0$ is the universal scalar
trilinear coupling and $m_{H_u}$ and $m_{H_d}$ are the soft masses for
the up-type and down-type Higgs, respectively. Scalar masses of MSSM
matter multiplets $(Q, \bar U, \bar D, L, \bar E)$ are taken to be
universal:
\begin{eqnarray}
m_Q = m_{\bar U} = m_{\bar D} = m_L = m_{\bar E} = m_0 \ {\rm at} \ M_{\rm inp}, \label{eq:cond2}
\end{eqnarray}
where we have omitted flavor indices.  We choose $m_{H_u}=m_{H_d}=0$
rather than $m_{H_u}=m_{H_d}=m_0$ in order to avoid the region with
unsuccessful electroweak symmetry breaking (EWSB); if
$m_{H_u}=m_{H_d}=m_0$ and $m_0 \gg M_{1/2}$, the EWSB does not
occurs~\cite{Feng:1999mn}.  Even if $m_{H_u}$ and $m_{H_d}$ are
non-vanishing, the bound on the gluino mass is almost unchanged in
most of the parameter space.\footnote{ The exception is the case where
  the EWSB occurs with the small $\mu$-parameter of a few hundred GeV,
  which will be discussed in Sec.~\ref{sec:dm}.}

In Fig.~\ref{fig:mh_mssm}, we show contours of the lightest Higgs
boson mass, $m_h$, on $\tan\beta$-$m_{\tilde g}$ plane, where
$m_{\tilde g}$ is a physical gluino mass. The Higgs boson mass is
computed by using {\tt FeynHiggs 2.11.3}~\cite{feynhiggs, feynhiggs2,
  feynhiggs3, feynhiggs4, feynhiggs5}. The mass spectrum of the SUSY
particles is calculated by using {\tt SuSpect 2.4.3}~\cite{suspect}.
The blue solid lines show $m_h$ of (127, 125, 123) GeV, from top to
bottom.  Although we expect the uncertainty in our calculation of the
Higgs mass of a few GeV, we use the contour of $m_h=125\ {\rm GeV}$ to
discuss how the existence of the extra matter fields affects the upper
bound on the gluino mass.  Then, in the MSSM, the upper-bound on the
gluino mass is as large as 7 TeV for $\tan\beta>10$ and $A_0=0$.\footnote
{ In the region $\tan\beta \simeq 1$, the gluino mass bound is as high
  as $10^{10}$ GeV~\cite{Bagnaschi:2014rsa}.  }
Such a heavy gluino is hardly observed by the LHC experiment.  In
addition, if the gluino is so heavy, the squark masses are also
expected to be so large via the RG effects unless there is an
accidental cancellation. Thus, in the MSSM, the discovery of the
colored SUSY particles is challenging unless the trilinear coupling of
the stop is sizable at the boundary $M_{\rm inp}$.

\subsection{Gluino mass bound for $N_5=3$ and $4$}

\begin{figure}[!t]
\begin{center}
\includegraphics[scale=1.05]{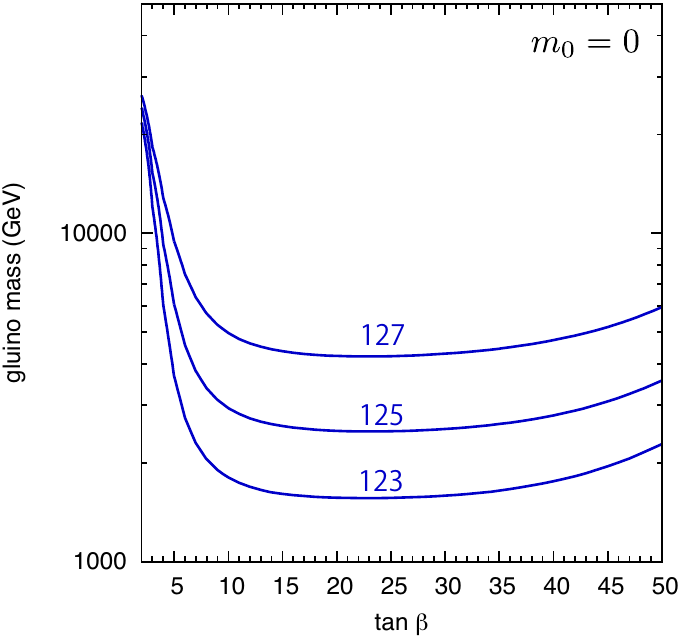}
\includegraphics[scale=1.05]{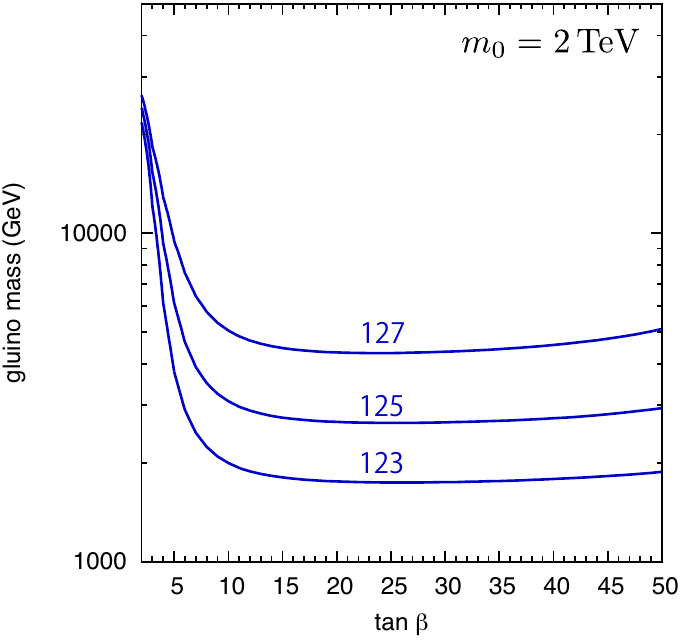}
\includegraphics[scale=1.05]{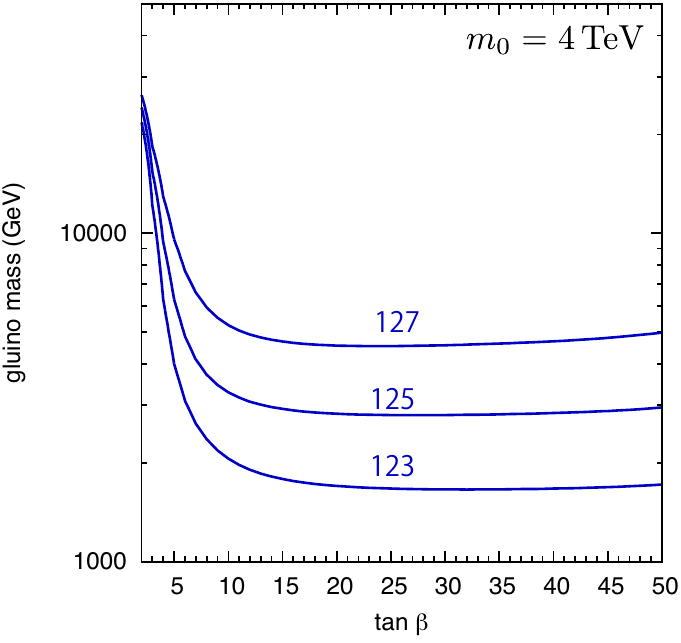}
\includegraphics[scale=1.05]{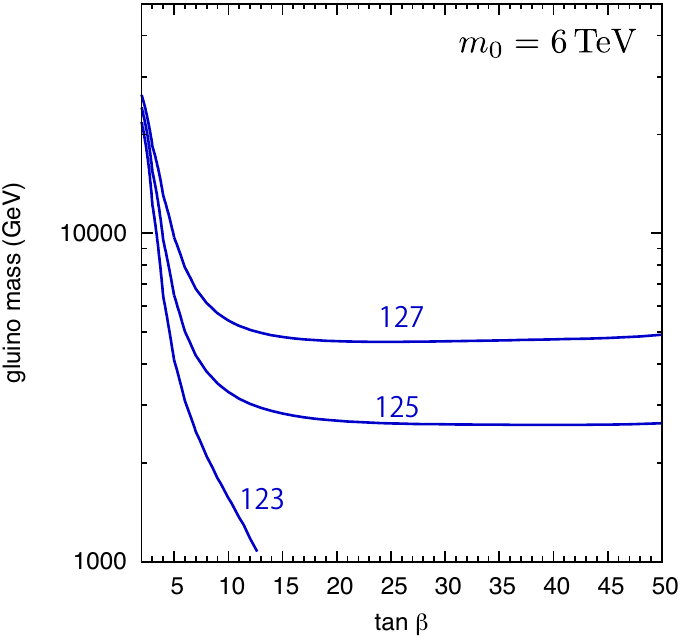}
\qquad
\caption{
Contours of $m_h$ on $\tan\beta$-$m_{\tilde g}$ plane for different $m_0$.
We take $N_5=3$ and $M_V=1$\,TeV. 
The solid lines correspond to $m_h=(127,125,123)$\,GeV from top to bottom.
}
\label{fig:mh_n53}
\end{center}
\end{figure}
%%%%%%%%%

\begin{figure}[!t]
\begin{center}
\includegraphics[scale=1.05]{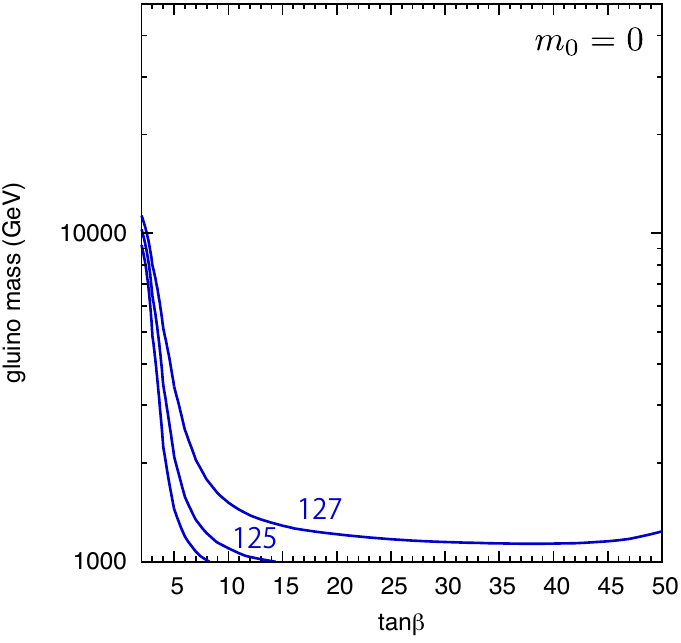}
\includegraphics[scale=1.05]{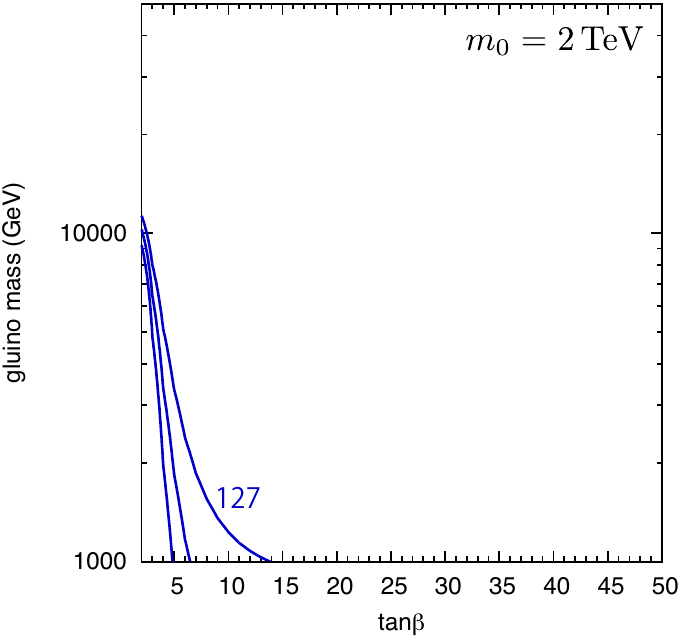}
\includegraphics[scale=1.05]{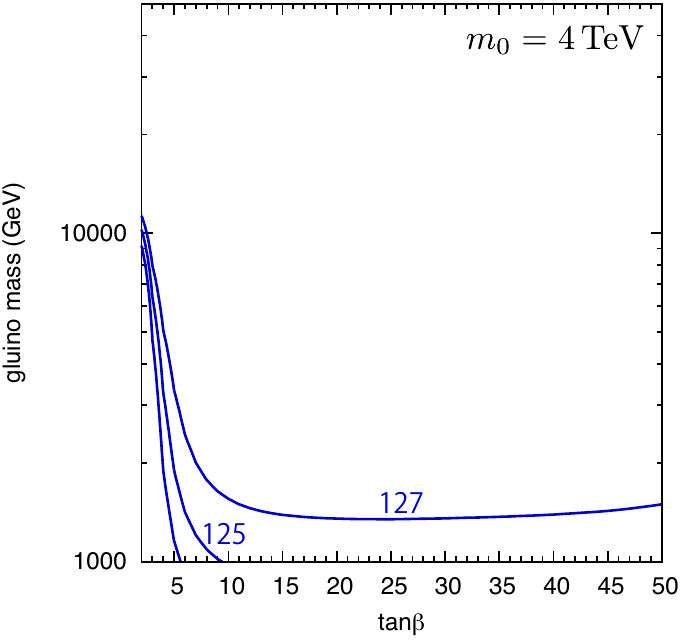}
\includegraphics[scale=1.05]{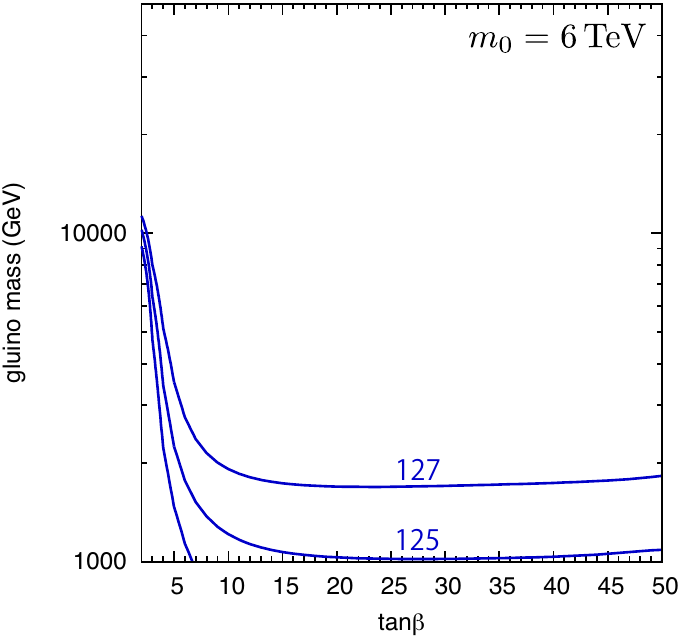}
\qquad
\caption{
Contours of $m_h$ on $\tan\beta$-$m_{\tilde g}$ plane for different $m_0$.
The solid lines correspond to $m_h=(127,125,123)$\,GeV from top to bottom.
Here, $N_5=4$. 
}
\label{fig:mh_n54}
\end{center}
\end{figure}
%%%%%%%%%

%%%%%%%%%%%%%
\begin{figure}[!t]
\begin{center}
\includegraphics{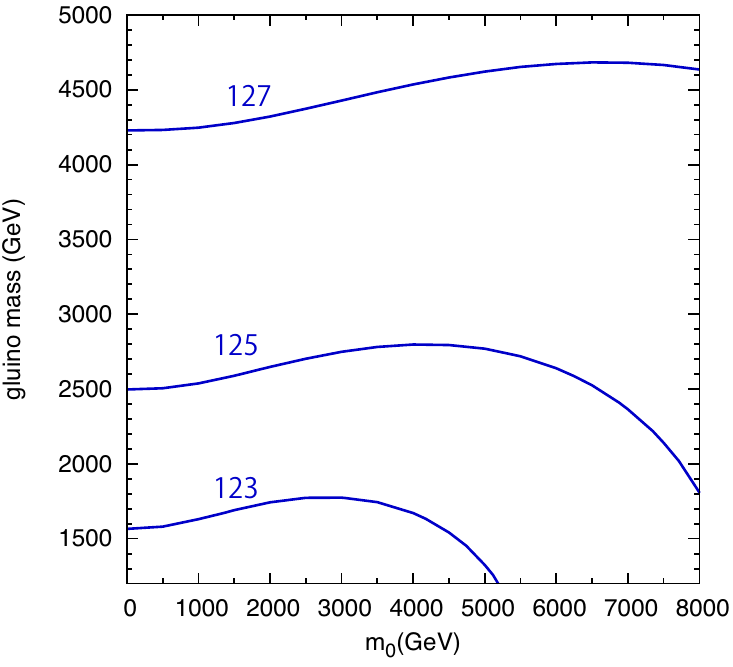}
\includegraphics{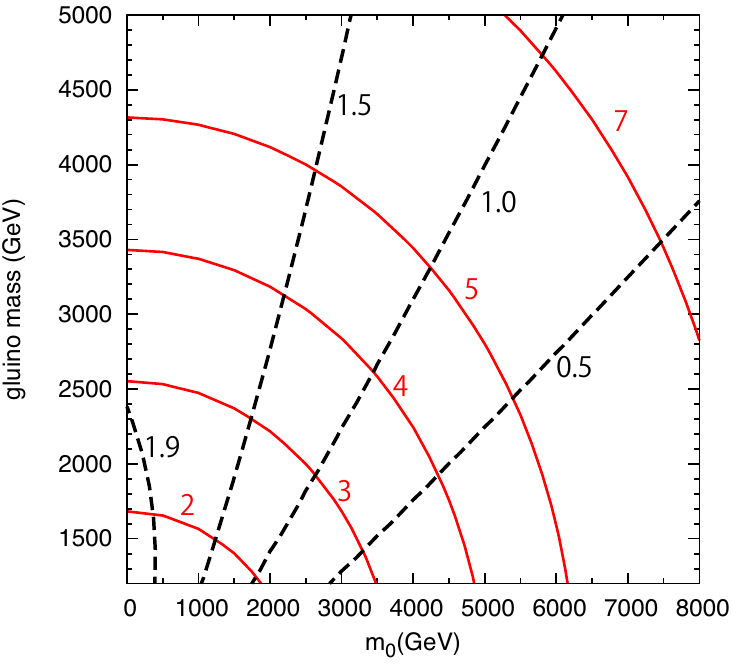}
\qquad
\caption{
Contours of $m_h$ in unit of GeV (left), $m_{\tilde t}$ and $|A_t^2|/m_{\tilde t}^2$ (right) for $N_5=3$.
The stop mass in unit of TeV and $|A_t^2|/m_{\tilde t}^2$ are shown in the red solid line and black dashed line, respectively. Here, $\tan\beta=25$.
}
\label{fig:n53_tb25}
\end{center}
\end{figure}
%%%%%%%%%

%%%%%%%%%%%%%
\begin{figure}[!t]
\begin{center}
\includegraphics{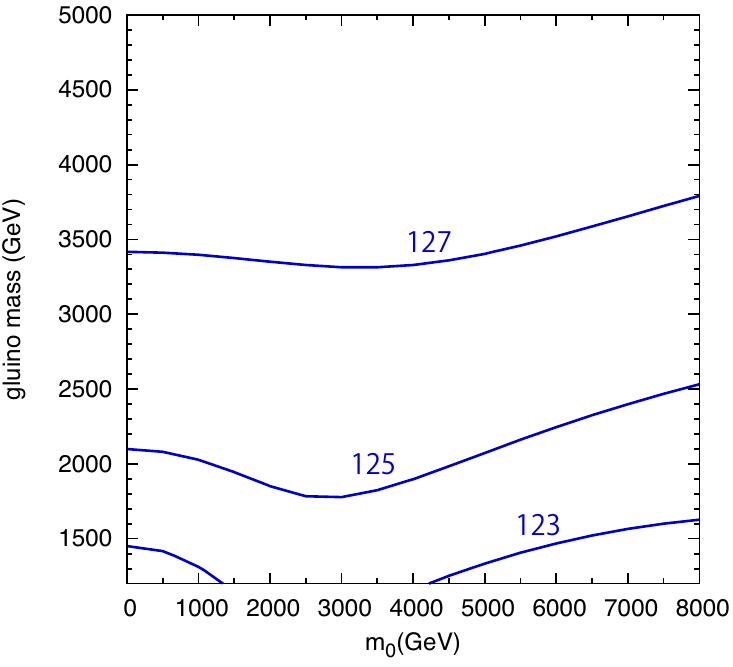}
\includegraphics{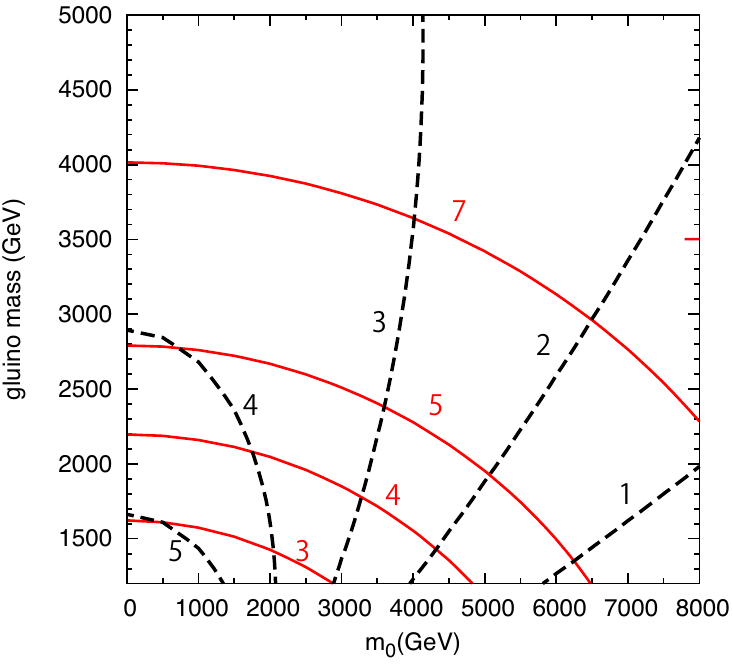}
\qquad
\caption{
Contours of the $m_h$ (left), $m_{\tilde t}$ and $|A_t^2|/m_{\tilde t}^2$ (right) for $N_5=4$. Here, $\tan\beta=5$.
}
\label{fig:n54_tb5}
\end{center}
\end{figure}
%%%%%%%%%

\begin{figure}[!t]
\begin{center}
\includegraphics[scale=1.05]{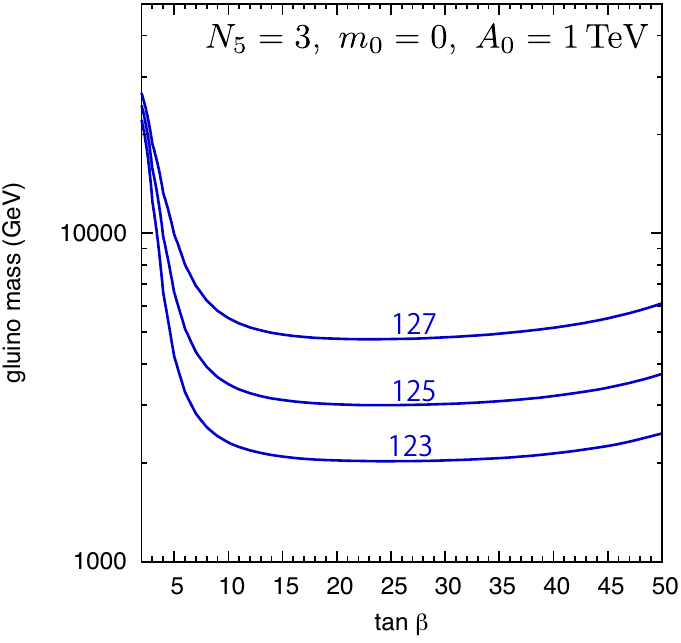}
\includegraphics[scale=1.05]{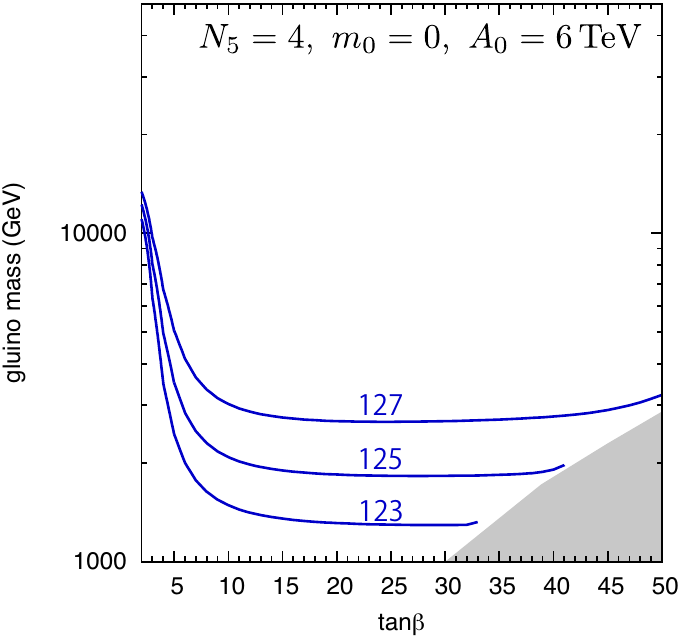}
\qquad
\caption{
The Higgs boson mass with non-zero $A_0$. In the left (right) panel, we take $N_5=3 (4)$ and $A_0=1(6)$\,TeV.
The gray region in the right panel is excluded due to the tachyonic stau.
}
\label{fig:mh_with_aterm}
\end{center}
\end{figure}
%%%%%%%%%

Next we show the results for $N_5=3$ and $4$, for which the upper
bound on the gluino mass is expected to be lower than the MSSM case.
In addition to the boundary conditions Eqs.~(\ref{eq:cond1}) and
(\ref{eq:cond2}), we take the scalar masses for the vector-like
multiplets to be universal:
\begin{eqnarray}
m_{\bar D'_i}=m_{D'_i}=m_{\bar L'_i}=m_{L'_i}=m_0 \ {\rm at} \  M_{\rm inp} \,.
\end{eqnarray}
The SUSY mass for the vector-like multiplets is taken to be
$M_V=1$\,TeV.  SUSY mass spectra are calculated by solving two-loop
RGEs with contributions from the vector-like multiplets and by
including one-loop threshold corrections to the gauge coupling
constants. These effects are included by modifying the {\tt SuSpect}
code.  The one-loop threshold corrections from the vector-like
multiplets are included by shifting the gauge couplings constants at
the SUSY mass scale $m_{\mathcal{S}}$:
\begin{eqnarray}
g_1^{-2}(m_{\mathcal{S}}) &\to& g_1^{-2}(m_{\mathcal{S}})
%- \frac{N_5}{8\pi^2} \frac{2}{3} 
%\ln \frac{m_{\mathcal{S}}}{M_V} 
%-  \frac{N_5}{8\pi^2} \frac{2}{5} 
%\left[ \frac{1}{4} \ln \frac{m_{\mathcal{S}}^2}{m_{L'_-} m_{L'_+}} 
%+ \frac{1}{6} \ln \frac{m_{\mathcal{S}}^2}{m_{D'_-}m_{D'_+}}  \right]
- \frac{N_5}{8\pi^2}
\left[
\frac{2}{3} 
\ln \frac{m_{\mathcal{S}}}{M_V} 
+ \frac{1}{10} \ln \frac{m_{\mathcal{S}}^2}{m_{L'_-} m_{L'_+}} 
+ \frac{1}{15} \ln \frac{m_{\mathcal{S}}^2}{m_{D'_-}m_{D'_+}}  \right]
\, ,\nonumber \\
g_2^{-2}(m_{\mathcal{S}}) &\to& g_2^{-2}(m_{\mathcal{S}}) - \frac{N_5}{8\pi^2}  
\left[ \frac{2}{3} \ln \frac{m_{\mathcal{S}}}{M_V} + \frac{1}{6} \ln \frac{m_{\mathcal{S}}^2}{m_{L'_-}m_{L'_+}} 
 \right] \, ,\nonumber \\
g_3^{-2}(m_{\mathcal{S}}) &\to& g_3^{-2}(m_{\mathcal{S}}) - \frac{N_5}{8\pi^2} 
\left[ \frac{2}{3} \ln \frac{m_{\mathcal{S}}}{M_V} + \frac{1}{6} \ln \frac{m_{\mathcal{S}}^2}{m_{D'_-}m_{D'_+}}
 \right] \, , \label{eq:thc}
\end{eqnarray}
where $m_{L'_{\pm}}$ ($m_{D'_\pm}$) are mass eigenvalues of the scalar components of $L_i'$ and $\bar L_i'$ ($\bar D_i'$ and $D'_i$).

In Figs.~\ref{fig:mh_n53} and \ref{fig:mh_n54}, we show contours of
$m_h$ on $\tan\beta$-$m_{\tilde g}$ plane in the presence of
vector-like multiplets for $N_5=3$ and $4$, respectively.  The scalar
mass $m_0$ is taken to be $m_0=0$, $2$, $4$, and $6$ TeV.  In the case
$N_5=3$, $m_h=125$\,GeV is realized with the gluino mass less than 3
TeV for large enough $\tan\beta$ (i.e., $\tan\beta\gtrsim 10$).  The
gluino with such a mass is expected to be detectable with the high
luminosity LHC~\cite{lhc_highlumi}.  In the case $N_5=4$, the
upper-bound on the gluino mass is even smaller; the gluino should be
lighter than $\sim 2.5$ TeV for $\tan\beta>5$.

In Figs.~\ref{fig:n53_tb25} and \ref{fig:n54_tb5}, we show contours of
$m_h$ (left), $m_{\tilde t}$ and $|A_t/m_{\tilde t}|^2$ (right) on
$m_0$-$m_{\tilde g}$ plane, where $m_{\tilde t}\equiv \sqrt{m_{Q_3}
  m_{{\bar U}_3}}$.  In Fig.~\ref{fig:n53_tb25}, we take $N_5=3$ and
$\tan\beta=25$.  The gluino mass is smaller than 2.8 TeV for
$m_h=125$\,GeV, because of sizable $|A_t/m_{\tilde t}|^2$, which
enhance the threshold correction to the Higgs mass, or relatively
large $m_{\tilde t} > 5$\,TeV.  In Fig.~\ref{fig:n54_tb5}, we take
$N_5=4$ and $\tan\beta=5$. The gluino mass is smaller than 2.5 TeV due
to the large $|A_t/m_{\tilde t}|^2$, which is as large as 4--5 for
$m_0 < 2$\,TeV.  We also comment on the $m_0$-dependence of the bound
on the gluino mass.  When $m_0$ is relatively small, the stop masses
are determined mostly by the gluino mass via the RG effects.  In such
a case, the upper bound on the gluino mass is insensitive to $m_0$.  On
the contrary, when $m_0$ is large, the stop masses becomes sensitive
to $m_0$; in such a case, with the increase of $m_0$, the upper bound
on the gluino mass becomes lower.

\subsection{The effect  of the bare $A$-term}\label{sec:aparam}

Let us discuss the effects of the non-zero $A_0$.  The bare
$A$-parameter, $A_0$, contributes to $A_t$ destructively
(constructively) if $M_{1/2}$ and $A_0$ have same (opposite) signs
(see Eq.~(\ref{At(fit)})).  Thus, with taking negative $A_0$,
$A_t(m_{\mathcal{S}})$ increase, and the gluino mass which realizes
$m_h\simeq125$\,GeV becomes smaller.  On the contrary, when $A_0$ is
positive and large, $A_t(m_{\mathcal{S}})$ becomes suppressed so that
the upper bound on the gluino mass can become higher.  Notice that the
effects of non-vanishing $A_0$ on the stop masses are not so
significant unless $|A_0|$ is very large (see Eqs.~(\ref{mQ3(fit)})
and (\ref{mU3(fit)})).

In Fig.~\ref{fig:mh_with_aterm}, we show the upper bound on the gluino
mass, taking non-vanishing $A_0 >0$.  In the case $N_5=3$ and
$A_0=1$\,TeV, the gluino mass bound slightly increases compared to the
case of $A_0=0$.  The gluino mass bound remains $\sim 3\ {\rm TeV}$
for $15 < \tan\beta < 35$.  In the case $N_5=4$, even if $A_0=6\ {\rm
  TeV}$, the gluino mass bound is as small as 2 TeV for
$\tan\beta>10$.

\subsection{Implications of dark matter} \label{sec:dm}

In the simple set up discussed above, the Bino-like neutralino is the
lightest SUSY particle (LSP) and its relic density is much larger than
the observed dark matter density, $\Omega_{\rm CDM} h^2 \simeq
0.12$~\cite{dm_density}; in such a case, the thermal relic lightest
neutralino is not a viable dark matter candidate.  However, with a
slight modification, the lightest neutralino can be dark matter
without much affecting the gluino mass bound.  Below, we discuss
several possibilities.

\begin{table*}[!t]
\caption{\small Mass spectra in sample points. We take $A_0=0$ and $M_{\rm inp}=10^{16}$\,GeV. 
Here, $A_t$ shown in the table is the generated $A$-term at $m_{\mathcal{S}}$.
}
\label{tab:sample}
%\begin{table}[]
\begin{center}
\begin{tabular}{|c||c|c|c|c|c|}
\hline
Parameters & Point {\bf I} & Point {\bf II}  & Point {\bf III} & Point {\bf IV} & Point {\bf V}\\
\hline
$N_5$  & 3  & 3                                & 4      & 4  &3\\
$M_{3}$\,(GeV) & 3000  & 3540      & 6900 & 6300 & 3400\\
$M_{1}/M_3$  & 1  & 1.0                  & 0.83 & 1 & 0.7\\
$M_{2}/M_3$  & 1  & 0.61                &  0.62 & 1 & 1\\
$m_0$\,(GeV)  & 0  & 0                    &  0 & 4000 & 0\\
$m_{H_{u,d}}$/1\,TeV & 3.441  & 0  &  0 & 6.392 & 0\\
$\tan\beta$ & 10 & 25                       & 6 & 5 & 32.9\\
\hline
$\mu$ (GeV) & 229 & 3410              & 5270 & 194 & 3210\\
$A_t$ (GeV) & $-4030$ & $-4450$       & $-7050$ & $-6720$ & $-4480$\\
\hline
%\hline
%
%
Particles & Mass (GeV) & Mass (GeV) & Mass (GeV) & Mass (GeV) & Mass (GeV) \\
\hline
$\tilde{g}$ & 2470 & 2970              & 1890  & 1760  & 2840\\
$\tilde{q}$ & 3670--3890 & 4340--4400 & 5360--5370 & 5900--6070 & 4150--4410\\
$\tilde{t}_{2,1}$ & 3220, 2130 & 3780, 3250 & 4390, 3130 & 4760, 2560 & 3720, 2940\\
$\tilde{\chi}_{2,1}^\pm$ & 942, 232 & 3410, 669  & 5240, 537 & 884, 196 & 3210, 1110\\
$\tilde{\chi}_4^0$ & 942 & 3410 & 5240 & 884 & 3210\\
$\tilde{\chi}_3^0$ & 537 & 3410 & 5240 & 590 & 3210\\
$\tilde{\chi}_2^0$ & 238 & 669 & 537  & 203 & 1110\\
$\tilde{\chi}_1^0$ & 227 & 645  & 510 & 191 & 420 \\
$\tilde{e}_{L, R}(\tilde{\mu}_{L, R})$ & 1510, 911 & 1140, 1080  & 1630, 1420 & 4580, 4260 & 1700, 715\\
$\tilde{\tau}_{2,1}$ & 1500, 860 & 1150, 983 & 1630, 1420 & 4580, 4250 & 1650, 423\\
$H^\pm$ & 3730 & 3290 & 5590 & 6910 & 3040\\
$h_{\rm SM\mathchar`-like}$ & 125.2 &  125.2  & 126.3 & 125.6 & 125.2\\
\hline
\end{tabular}
%\end{table}
\end{center}
\end{table*}

\begin{itemize}
\item Higgsino dark matter

  If $m_{H_u}^2(M_{\rm inp})$ is tuned, the EWSB can occur with $\mu
  \sim m_Z$.  In such a case, the Higgsino of a few hundred GeV can be
  the LSP.  The thermal relic abundance of such Higgsino LSP is
  smaller by $\sim 1/10$ compared to the observed dark matter
  abundance.  However, the observed dark matter abundance can be
  realized with non-thermal productions~\cite{Giudice:1998xp,
    Moroi:1999zb,Gelmini:2006pw}.  With smaller value of the Higgsino
  mass, the Higgs boson mass is raised about 1\,GeV for the fixed
  gluino mass, and hence the bound on the gluino mass becomes
  stronger.

\item Bino-Wino coannihilation

  If we relax the GUT relation among the gaugino masses, the mass
  difference between the Bino and Wino can be small and the thermal
  relic abundance of the lightest neutralino can be consistent with
  the observed dark matter density, because of the Bino-Wino
  coannihilation with $M_2/M_1(M_{\rm inp}) \sim
  0.5-0.7$.  This reduces the Higgs boson mass only
  slightly.  Thus, the upper-bound on the gluino mass is almost
  unchanged.

\item Wino dark matter

  If the mass ratio of $M_2/M_1$ is even smaller than the previous
  case, the lightest neutralino can be (almost) Wino-like.  Although,
  the thermal relic abundance of the Wino-like neutralino is too small
  as in the case of the Higgsino dark matter, with non-thermal
  production, the relic abundance can be consistent with the observed
  dark matter relic.  The constraint from gamma rays from dwarf
  spheroidal galaxies gives a lower-bound on the Wino mass,
  $M_2(m_\mathcal{S}) \gtrsim 320$\,GeV~\cite{wino_indirect}.  The LHC
  may discover/exclude the Wino LSP with the mass up to $\sim 500\
  {\rm GeV}$ through electroweak productions~\cite{wino_lhc}.

\item Bino-stau coannihilation

  For $N_5=3$ and large $\tan\beta$, the mass difference between stau
  and neutralino becomes small and the relic abundance of the
  neutralino is reduced due to the Bino-stau
  coannihilation~\cite{Ellis:1998kh, Ellis:1999mm}.  Smaller $M_1$ at
  $M_{\rm inp}$ helps to reduce the mass difference.

\end{itemize}

In the Table~\ref{tab:sample}, we show sample points where the relic
density of the lightest neutralino is consistent with the observed
dark matter density. We calculate the thermal relic density using {\tt
  MicrOMEGAs 4.1.7} package~\cite{Belanger:2001fz, Belanger:2004yn}.
In the points {\bf I}, {\bf II} and {\bf V} ({\bf III} and {\bf IV}),
we take $N_5=3$ ($N_5=4$).  In the points {\bf II}, {\bf III} and {\bf
  V}, the GUT relation among the gaugino masses is relaxed.  In {\bf
  II} and {\bf III} ({\bf V}), because of the Bino-Wino coannihilation
(stau coannihilation), the thermal relic abundance of the lightest
neutralino becomes constraint from the observed dark mater abundance.
In the points {\bf I} and {\bf IV}, the Higgsino like neutralino is
the LSP, and the thermal relic abundance smaller by about 1/10
compared to the observed dark matter abundance. With non-thermal
productions, this Higgsino-like neutralino can be candidate for a dark
matter.\footnote
{One might worry about the constraint from the direct detection
  experiments of dark matter.  However, the neutralino-nucleon
  scattering cross section is suppressed when $M_1$ is large.
}

\section{Conclusion and discussion}

We have investigated the upper-bound on the gluino mass for $m_h
\simeq 125$\,GeV, with 3 or 4 copies of the vector-like multiples
around TeV, transforming ${\bf 5}$ and $\bar{\bf 5}$ representations
in $SU(5)$ GUT gauge group.  We have shown that with these vector-like
multiplets, the upper-bound on the gluino mass is significantly
reduced compared to that of the MSSM.  The significant reduction
originates from the fact that the radiatively generated trilinear
coupling of stops as well as the stop masses is enhanced for the fixed
gluino mass at the low-energy scale.  In both cases of $N_5=3$ and
$4$, the gluino mass is less than 3\,TeV in a wide range of parameter
space, and the gluino is likely to be discovered at the LHC Run-2 or
the high luminosity LHC.

The presence of 3 or 4 copies of the vector-like multiplets at around
TeV is suggested by the recently observed 750\,GeV diphoton excess at
the LHC.  With the vector-like extra matters as well as a gauge
singlet field $\Phi$, the diphoton excess can be explained by
the production and the diphoton decay of $\Phi$.  In such a scenario, the
singlet field $\Phi$ should couple to the vector-like matters ${\bf 5}$
and $\bar{\bf 5}$ with Yukawa couplings of $\sim 1$ in the
superpotential.  In order to make the mass of $\Phi$ as well as those of
the vector-like matters being close to the TeV scale, it is attractive
if they have the same origin, i.e., the vacuum expectation value of
$\Phi$, $\left<\Phi\right> \simeq$ TeV.  Even if the SUSY breaking soft mass
squared of $\Phi$, denoted as $m_\Phi^2$, is positive at a high scale, it
can be driven to negative via the RG effect because $\Phi$ is expected to
have a strong Yukawa interaction with the vector-like matters.  With
such a negative mass squared parameter, $\left<\Phi\right>$ can become
non-vanishing.  However, we found that $|m_\Phi^2|$ is typically much
larger than $(1\,{\rm TeV})^2$ in the parameter region of our
interest.  Thus, we need a tuning to obtain the correct size of
$\left<\Phi\right>$ by introducing supersymmetric mass parameters for
$\Phi$:\footnote{
Even in a case that the masses of the vector-like matters do not originate from the non-zero $\left<\Phi\right>$, 
in order to make a mass of a scalar component of $\Phi$ to be 750\,GeV, one generally requires a tuning due to the negative $m_\Phi^2$.
}
the required tuning is typically $\mathcal{O}(1)$ percent level.

%%%%%%%%%%%%%%%%%%%%%%%%%%%%%%%%%%%%%%%%%%%%
\section*{Acknowledgments}
%%%%%%%%%%%%%%%%%%%%%%%%%%%%%%%%%%%%%%%%%%%%
This work is supported by JSPS KAKENHI Grant Numbers 
JP26400239 (T.M.), JP26104009 (T.T.Y),
JP26287039 (T.T.Y.), JP16H02176 (T.T.Y),
JP15H05889 (N.Y.) and JP15K21733 (N.Y.);
and by World Premier International Research Center Initiative (WPI Initiative), MEXT, Japan (T.M and T.T.Y.).

\end{document}